\documentclass[onecolumn,showpacs,preprintnumbers,amsmath,amssymb]{revtex4}
\usepackage{graphicx}

\begin{document}
\title{Dynamical failure of Turing patterns}
\author{Alon Manor and Nadav M. Shnerb}
\affiliation{
  Department of Physics, Bar-Ilan University, Ramat-Gan
52900 Israel }

\begin{abstract}
The emergence of stable disordered patterns in reactive systems on a
spatially homogenous substrate is studied in the context of
vegetation patterns in the semi-arid climatic zone. It is shown that
reaction-diffusion systems that allow for Turing instability may
exhibit heterogeneous "glassy" steady state with no characteristic
wavelength if the diffusion rate associated with the slow reactant
is very small.  Upon decreasing the diffusion constant of the slow
reactant three phases are identified: strong diffusion yields a
stable homogenous phase, intermediate diffusion supports Turing
(crystal like)  patterns while  in the slow diffusion limit the
glassy state is  the generic stable solution. In this disordered
phase the dynamics is of crucial importance, with strong differences
between local and global initiation.
\end{abstract}

\maketitle
\section{Introduction}
Pattern formation in reactive systems is a well investigated
research field since the pioneering work of Turing \cite{turing}
half a century ago. The idea of diffusion induced instability has
been found to be applicable in many systems, ranging from animals
coat patterns  \cite{murray} to chemical reactions \cite{chem} and
was analyzed mathematically using variety of techniques
\cite{cross}. Generically, the instability of the spatially
homogenous state is attributed to the existence of two different
diffusion rates associated with the different reactants, where the
inhibitor diffusion constant is faster than that of the activator.
The resulting patterns are crystal-like, with a typical wavelength
that, in many cases, is close to the characteristic length scale of
the linear instability.

Recently, the applicability of Turing's idea to vegetation patterns
in the semi-arid climatic zone has been studied by many authors
\cite{meron+}. The dynamics of this class of ecological systems is
governed by the competition of perennial biomass units (shrubs,
trees) for common resource (water). Taking into account some kind of
"positive feedback" mechanism (like the slower mobility of soil
moisture in the vicinity of the biomass), the corresponding
reaction-diffusion equations lead to a Turing-like instability that
may correspond to the observed ordered patterns.

\begin{figure}
  \includegraphics[width=10cm]{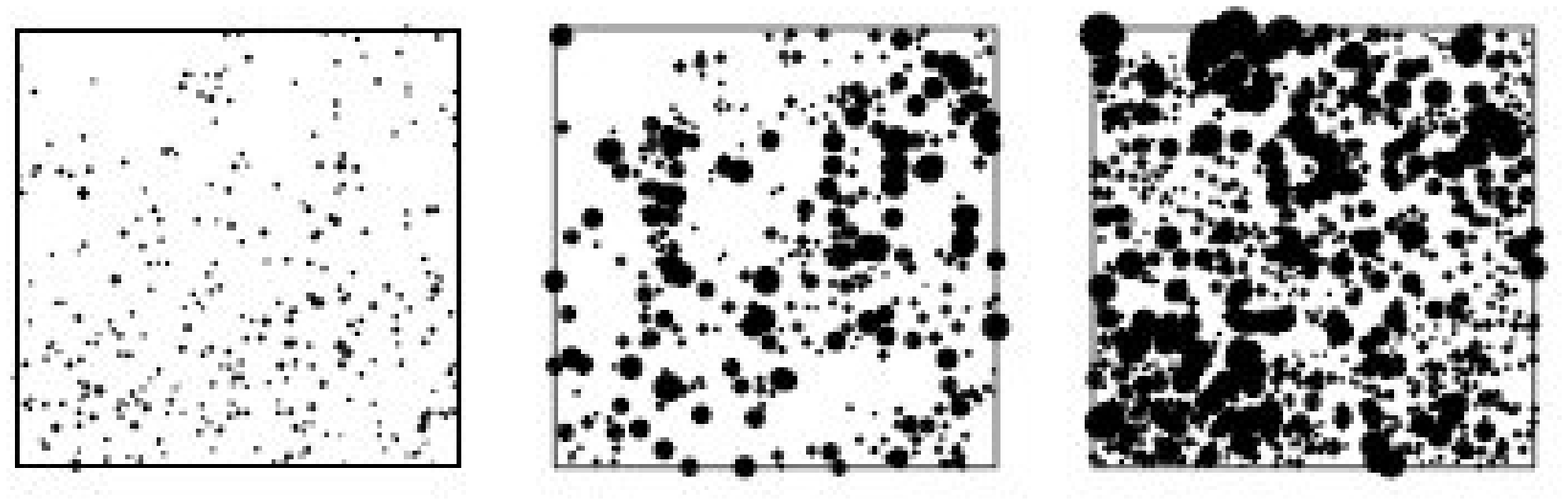}
  \caption{Results of direct field measurements at three different locations along the precipitation
gradient. The distribution of perennial shrubs (annual flora not
included) is presented for an area of 100 square meters at each
site. Each black spot presents a shrub and the size of a spot is
proportional to the area  of the canopy. Shrubs distribution on
hillslopes has been taken at three sites representing mildly arid,
semiarid, and subhumid climate conditions in Israel (see
\cite{lavee} for details).}
  \label{fig1}
\end{figure}

It should be noted, however, that the generic situation in the
semi-arid zone is \emph{disordered} vegetation patterns, as clearly
seen in Figure (\ref{fig1}), where the results of  typical field
observations  are presented. Interestingly, these patterns are quite
robust and may stay unchanged  for a long time (even hundreds of
years). This is a somewhat surprising feature, as one expects a
disordered system to wander among many microscopic configurations
with equivalent features. The best known examples of  disordered,
\emph{stable} patterns are observed when a supercooled  liquid fails
to reach its crystalline phase and freezes into a glass. Hence, the
term "glassy phase" or "reactive glass" is used here  to describe
this phenomenon, already known in the field of coupled lattice maps
\cite{clm}.

A non-Turing mechanism, based on the periodicity of the water supply
(dry and humid seasons) has already been suggested \cite{lavee} in
order to model the spontaneous spatial segregation into disordered
stable patterns. In this letter we intend to present another route
to the glassy state, a dynamical failure of the patterning at low
diffusion on a lattice \cite{lattice}. This mechanism seems to be a
generic property of many Turing systems; accordingly, the
water-vegetation equations are used here only to exemplify these
features and to gain some basic intuition about their origin. The
main aspect emphasized here is the appearance of many stable
spatially disordered solutions for nonlinear dynamics.  Applications
to (and discussion of) vegetation patterns will be presented
elsewhere.

Our basic observation is that the glassy phase exists, in many
autonomous Turing systems on a lattice, provided that the mobility
of the slow reactant is taken to be small enough. In that case
generic initial conditions flow into a stable disordered pattern,
long range spatial correlations disappear and the number of stable
solutions grows exponentially with the size of the system. The
deterministic reaction-diffusion dynamics is thus similar to the
behavior of a dissipative, low temperature glassy system, that flows
in phase space to the "closest" free energy local minima.

The importance of the dynamics is related to another analogy with
the physics of glasses.  The crystallization process in supercooled
liquids takes place via the growth of a single nucleus, and if the
growth rate is much slower than the production rate of microscopic
nuclei the system fails to display long range order. In our system
an analogous distinction should be made between global and local
initiation. Starting from a single localized "seed" the biomass
spreads into the unstable region  and  forms an ordered structure.
Global initiation from random state, on the other hand, yields
glassy pattern as small nuclei with different order parameters fail
to merge into a macroscopic Turing state. In the following the
glassy phase emergence from global initiation is first discussed,
while the ordered patterns associated with local initiation are
considered later on.

\section{Characteristics of the "glassy" phase}

Let us present a simple model for pattern formation in a
biomass-water system.  Denoting the water density by $w$ and the
biomass density by $b$, a simple, dimensionless set of
reaction-diffusion equations is:
\begin{equation}\label{dynamics}
\dot{w}={R-\lambda w b-w+D_w\nabla^2 w} \qquad \dot{b} = {w b -
\mu(b) b+D_b\nabla^2 b}
\end{equation}
 The first equation describes the water (or soil moisture)
 dynamics, with constant deposition (rain) $R$, inorganic losses
 due to percolation and evaporation ($-w$), consumption of water by the
 biomass ($\lambda w b$) and diffusion $D_w$. The second equation
 expresses the biomass time evolution: it decays
  at rate $\mu(b)$, grows upon water intake, and diffuses
 at a much smaller rate, $D_b<<D_w$.

\begin{figure}
 \includegraphics[width=13cm, angle = -90]{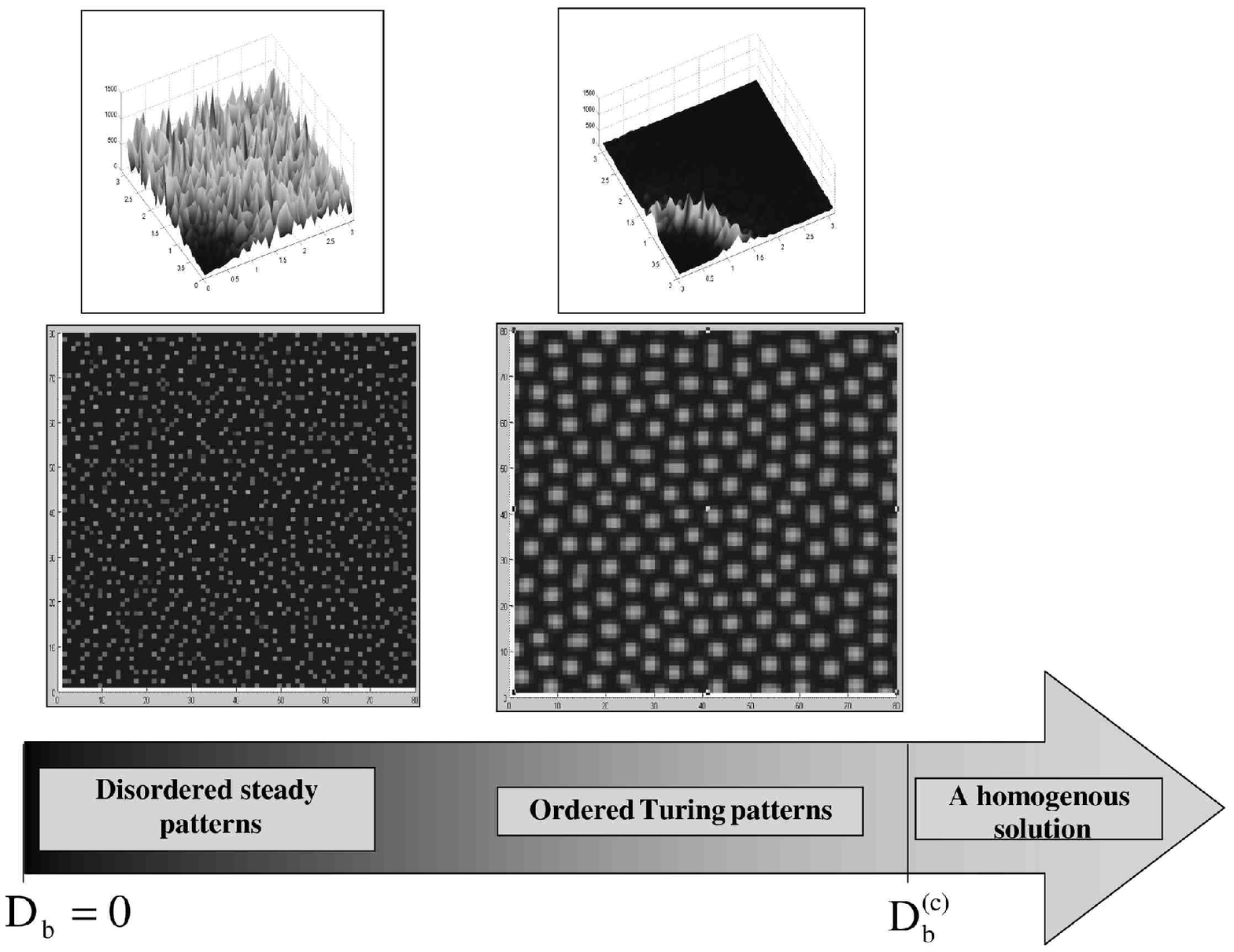}
  \caption{Spatial patterns (lower panels) and  their corresponding Fourier transform (upper panels)
  for different values of biomass diffusion.
  Numerical results of forward Euler
integration of the reaction-diffusion equations (1-2) on an $80
\times 80$  sites grid with periodic boundary conditions are
presented. In the right panel an ordered pattern is formed for
$D_b=0.02$ and in the left panel a disordered pattern for
$D_b=0.001$ . The simulation parameters are
    $\mu_0=0.1$, $ \mu_1=0.3$,  $R=0.5$, $D_w = 10$ and $\lambda = 1.2$.
    Initial conditions are no water and a seed of biomass taken  (at each site) from a
square distribution between [0,0.01].}\label{fig2}
\end{figure}

 This simple set of equations  supports  spatial
 segregation only if there is some "positive feedback" mechanism that
 facilitates growth (at a spatial location)
 in the presence of biomass.
 In fact, perennial flora  should
 reach some critical biomass threshold in order to survive the dry season,
 and  this feature may be implemented in smooth reaction-diffusion equations
 by assuming  smaller decay rate  for larger
 shrubs.  In this paper the following functional
 dependence  of the decay rate  $\mu$ is assumed,
\begin{equation}
\mu(b)={\mu_0+\frac{\mu_1 }{b+1}}\label{DeathRate}
\end{equation}
but the only essential trait is the monotonic decrease of the death
rate as a function of the biomass size. With (\ref{DeathRate}), Eq.
(\ref{dynamics}) admits a Turing instability, and below some
critical value of $D_b$ ordered patterns emerge,  as demonstrated in
the right panel of Figure \ref{fig2}.

What happens if $D_b$ is taken to be even smaller? According to the
standard linear analysis around the homogenous fixed point one
expects to see ordered patterns all the way down to $D_b = 0$, but
on a lattice one gets, instead, the result shown in the left panel
of Figure \ref{fig2}, with no apparent order or characteristic
length scale.  We stress that these disordered patterns are, indeed,
\emph{linearly stable}, as implied by direct numerical
diagonalization of the linearized evolution operator. Moreover, both
global initiation from the nearly  empty state (small "seeds" of
random height) and from the homogenous solution (small random
fluctuations around it) yield the same type of disordered steady
state.

\begin{figure}
 \includegraphics[width=10cm]{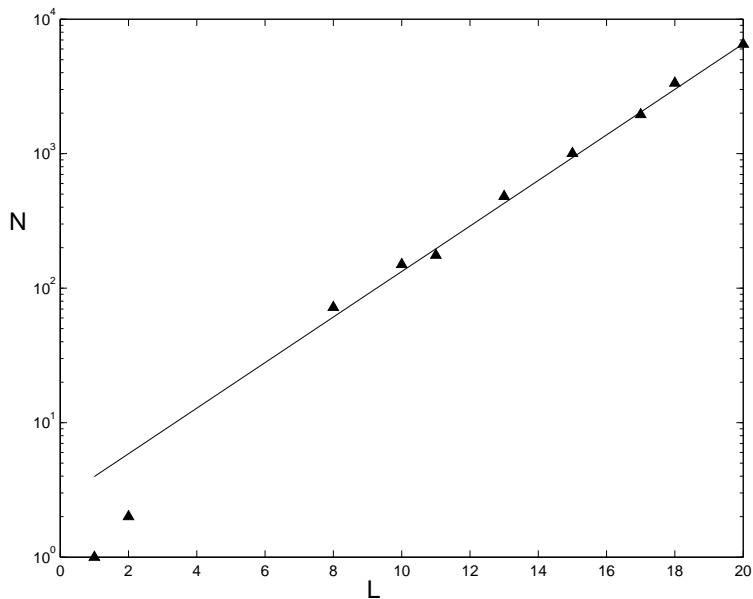}
  \caption{Numerical counting  of the number of stable solutions (N) as a function of the system
  size (L)
   for small, one dimensional samples where the  sizes ranging from
   1
  to  20 sites. Solutions were found numerically using Newton-Raphson method with random sampling of the
  initial conditions phase space and their stability was checked by linear analysis.
   The numerical results (triangles) fit to an exponential growth
  with system size, $N \sim exp(0.39 L)$, as indicated by the full line.
  (Boundary conditions are periodic, and all parameters
   are the same as in Fig. \ref{fig2}. }
  \label{fig4}
\end{figure}

One may guess, thus, that for small $D_b$ the phase space admits
many attractive fixed points, each corresponds to a different
disordered configuration. If a generic initial condition lays in the
basin of attraction of one of these points, the deterministic
evolution determined by Eq.  (\ref{dynamics}) simply takes the
system to that stable disordered point. Explicit counting of the
number of stable fixed points (for small systems) shows, in fact,
exponential growth with the system size in the glassy regime, as
indicated in Figure \ref{fig4}. This exponential growth reflects the
number of possible combinations of local patches and,
correspondingly,  the absence of long range order in the steady
solutions.

As $D_b$ decreases, thus, the system undergoes  an order-disorder
transition, characterized by the loss of long range order and by the
exponential dependence  of the  number of stable solution on the
system size.  A detailed analysis of that transition is beyond the
scope of this paper; here we present some qualitative results that
may be considered as a rough measure of that transition.

\begin{figure}
 \includegraphics [angle=270, width=18cm] {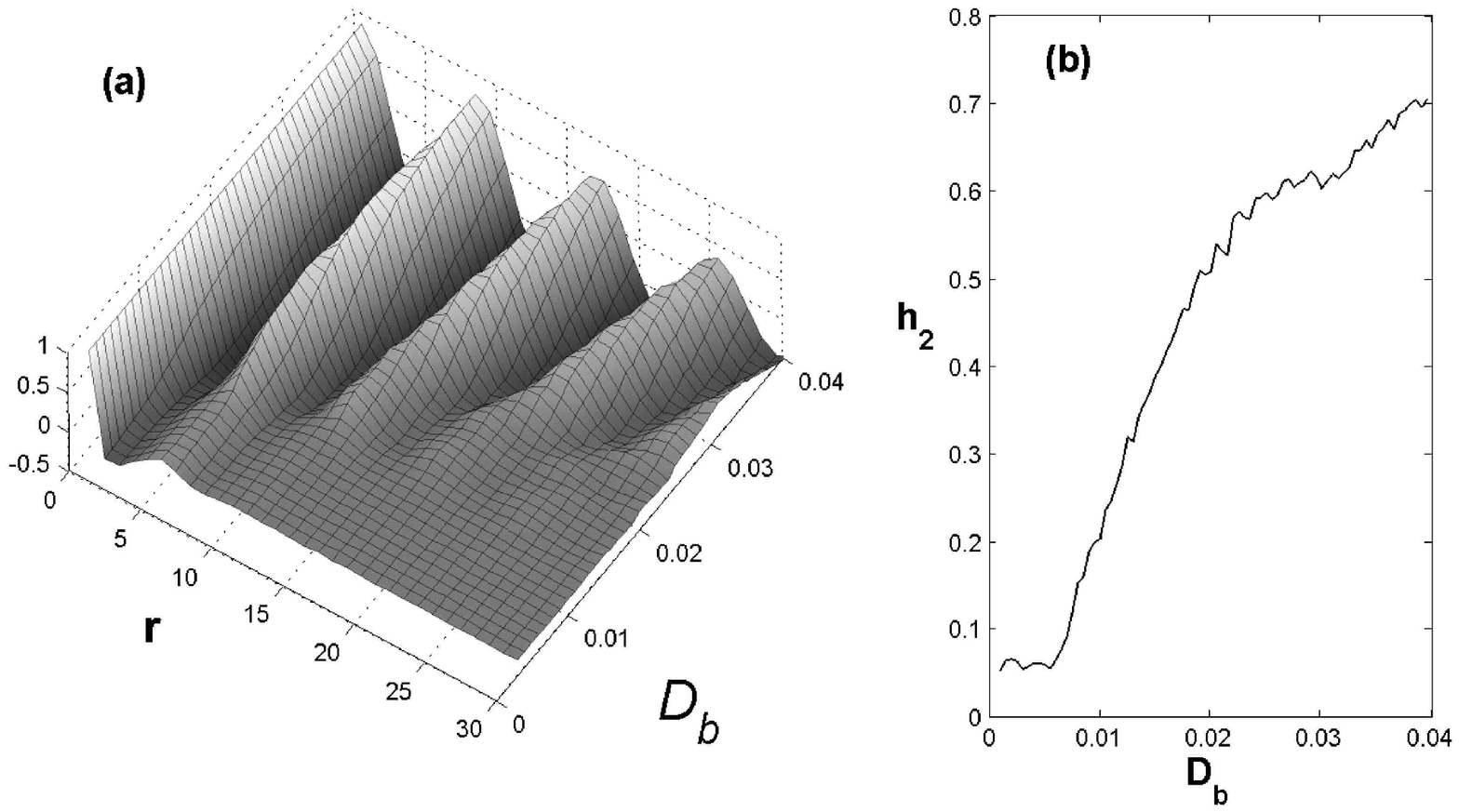}
  \caption{(a) The correlation function $g(r)$ (in one dimension) is plotted against $r$  for different values of $D_b$, where all other model
  parameters are the same as in Figure \ref{fig2}. Clearly, there is a decay of the spatial correlation function as $D_b$ approaches zero. In the right
  panel (b)
   the second peak's hight, $h_2$, is plotted against $D_b$. }
  \label{fig3}
\end{figure}

As a basic indication for the loss of long range order, the
normalized radial correlation function is plotted against the radius
as $D_b$ changes [Figure \ref{fig3}, left panel]. While for large
$D_b$ the correlations are  long range, at small diffusion only the
first peak (that corresponds to the effective competition length)
survives and there is almost no structure beyond that length scale.
The normalized height of the second peak of the radial correlation
function is  also plotted (right panel) against the biomass
diffusion coefficient and is shown to decrease with $D_b$ (almost
linearly) up to some saturation at low values. One may identify the
"critical" value of the diffusion constant with its value at the end
of the plateau, where the system starts to respond to the increase
of biomass spatial diffusion.

\begin{figure}
 \includegraphics[width=10cm]{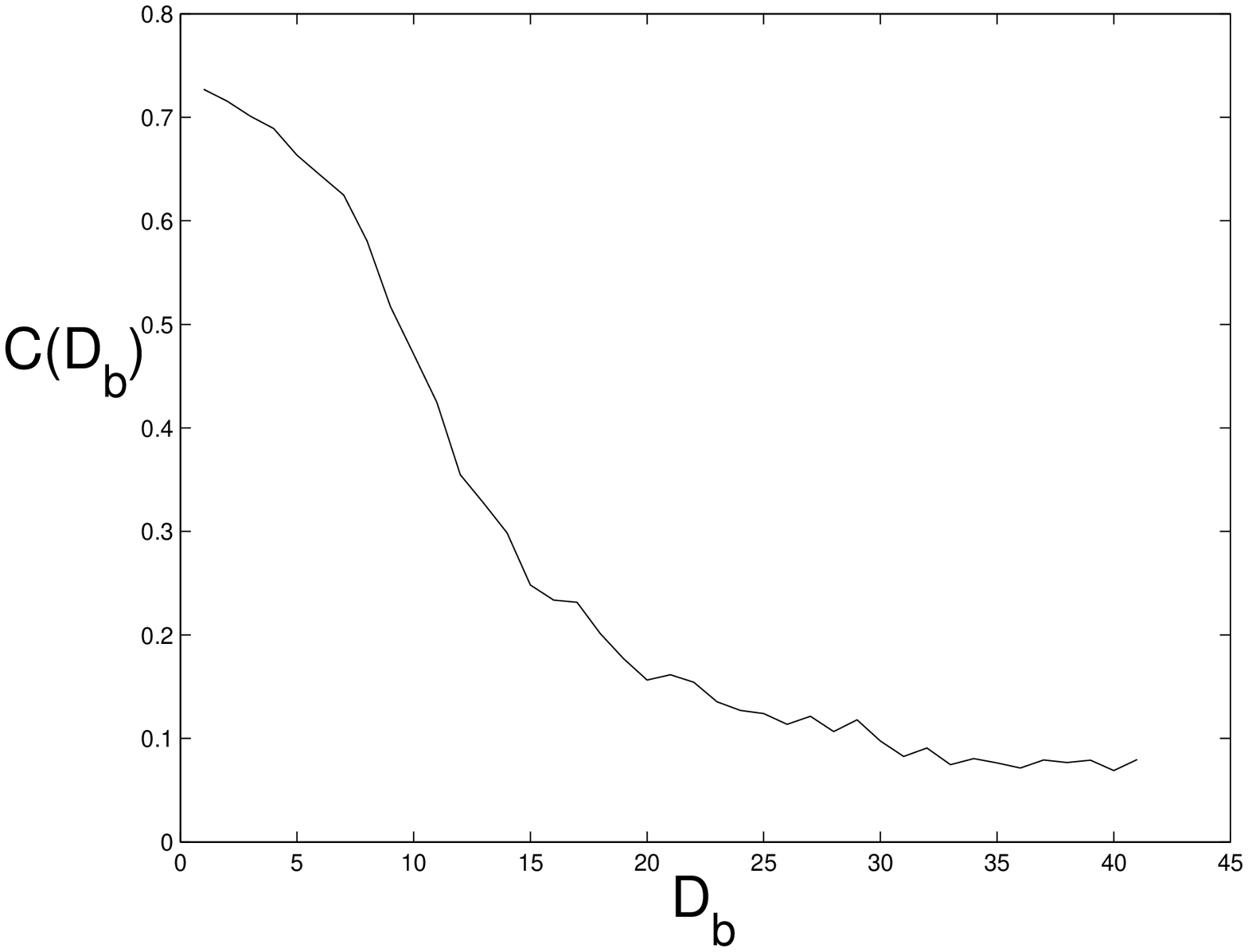}
  \caption{The overlap $C(D_b)$ between the initial conditions and the resulting pattern is presented for different $D_b$'s. With the same model
  parameters and initial conditions used above, the one dimensional steady state configuration shows much more overlap in the glassy phase than in the Turing phase.
  The initial conditions were mapped to a binary sequence (any local
  maxima is one, other points are zero) and the $L_2$ overlap of
  this vector with the final state is shown.}
  \label{overlap}
\end{figure}

Another manifestation of the transition is the overlap of the steady
state with the initial conditions. For a system dominated by a
single attractive fixed point there is a complete memory loss, as
all different initial states end up to be the same. As the number of
attractive points increases it is plausible to assume that the phase
space is divided into  non-overlapping, compact basins of
attraction, and a generic initiation  flows to its "own" fixed point
(in analogy with the diverging number of free energy local minima in
glassy systems). In Figure \ref{overlap}, the overlap  of the
initial conditions and the final solution is plotted against the
diffusion of the slow reactant, and the decrease of overlap as the
system approaches the ordered Turing limit is clearly seen.

\section{"Rich get richer" mechanism for disordered segregation}

Our numerics suggests that this crossover from ordered to glassy
patterns is a generic feature of lattice Turing systems:  the same
behavior has been observed for many different Turing systems we have
 examined (including the Gray-Scott  and the Gierer-Meinhardt
equations). Interestingly, our vegetation-water model also provides
us  with a simple intuitive argument that explains this transition.

The basic mechanism beyond the Turing instability  is the "rich get
richer" mechanism (Eq. \ref{DeathRate}) i.e., the fact that large
shrubs are less effected by water shortage, so that even if small
biomass units decay, large units still grow.

Let us consider the case of $N$ independent  biomass units using the
same water resource, namely, the limit $D_b \to 0$, $D_w \to
\infty$, of Eqs. (\ref{dynamics}). The  biomass growth leads to a
depletion of the water resource until, when $\mu(b)=w$ (for a
specific shrub) it  ceases to grow any more. However, at this
resource level larger biomass units continue their development and
the reduction of available resource, so the smaller shrubs start to
wilt. As this process continues, only the largest biomass unit (at
$t=0$) survives in the system. The mechanism of "rich get richer",
thus, amplify small differences of the initial state to the level of
"winner takes  all", where only one unit survives the competition
for common resource.

Turning back to the spatial model on a lattice with finite biomass
and water diffusion, our numerical simulations  indicate that the
same logic may hold for small $D_b$ and large $D_w$. One may
interpret these numerical results in view of the above argument:
finite (but large) water diffusion leads to spatial segregation of
the surface into large patches of competing biomass units, and the
same "winner takes  all" mechanism holds within each patch. For very
small biomass diffusion each of these patches is still dominated by
one large shrub, and the growth is so fast that the initial
conditions dictate the final state of the system. The situation is
close to the case of random sequential adsorption \cite{rsa}, where
the system fails dynamically to reach its optimal filling. Larger
biomass diffusion corresponds to  slower takeover process, and the
system dynamics admits stronger spatial correlations and allows for
the emergence of Turing pattern.

\section{Local initiation vs. global initiation}

\begin{figure}
  \includegraphics[width=10cm]{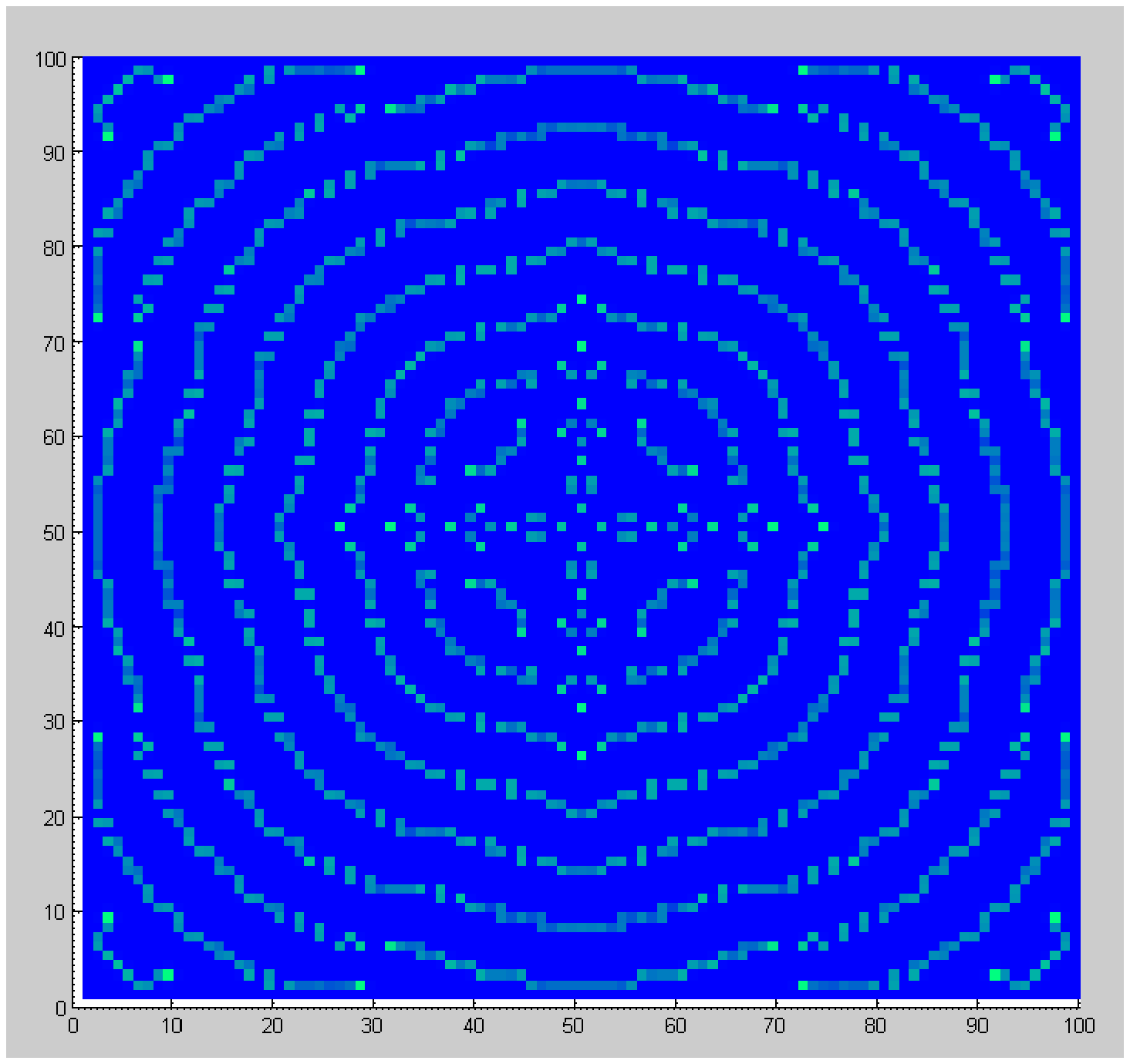}
  \caption{Numerical results for local initiation in the glassy
  regime. Here, the case $D_b = 0.002$ is evaluated where the
  initial conditions are only one seed in the middle of the
  sample, where all other parameters are the same as in Figure
  \ref{fig2}. The deviations from exact polar symmetry are due to
  the effect of the underlying square    lattice.}\label{fig5}
\end{figure}

As emphasized above, the failure of the system to produce a Turing
pattern is \emph{dynamical}, as many localized excitations fail to
merge into an ordered phase. This leads to the appearance  of strong
differences between different types of initial conditions. The
results presented in the last sections are typical to the case of
global initiation, where a homogeneous state is perturbed randomly
at each site. If, on the other hand, there is a single "grain" (one
perturbation with compact support) that grows and invades the
unstable region, the system flows to the ordered phase even if the
diffusion coefficient of the biomass is taken to be arbitrarily
small (and the invasion of the stable solution into the unstable
state becomes very slow). Comparing Figure (\ref{fig5}) and the low
$D_b$ pattern in Figure (\ref{fig2}), the differences between local
and global initiation are clearly recognized.

\section{Conclusions}

This paper deals with the possibility of dynamical failure in the
process of development of  Turing type patterns. This possibility
has been demonstrated for the case of water-vegetation system on a
lattice, where the biomass diffusion is small, such that its effect
fails to compensate the faster "rich get richer" mechanism. The
resulting patterns are disordered and robust without a
characteristic length scale, and are strongly correlated to the
random initial conditions. A Turing pattern does appear, for the
same set of parameters, if the system is initiated locally, as the
"rich get richer" ordering of shrubs according to their size
coincides with the spatial ordering.

Although the numerical experiments presented here are only for the
specific model of resource competition, the argument beyond the
dynamical failure is generic, and one expects to see similar
behavior in other Turing systems. For a system on finite lattice, at
least, it is plausible to expect different localized excitations
that (if the coupling between different  points is not strong
enough) fail to merge into one global pattern.

In analogy with  coupled lattice maps \cite{clm}, one may expect the
glassy phase to disappear at the continuum limit  (i.e., where the
lattice size approaches zero). Note, however, that in any reactive
system some degree of spatial discretization is dictated by the
discrete nature of the reactants. In our case, for example, the
discrete nature of a single shrub yields a basic length scale  that
corresponds to the minimal distance between the mother shrub and its
propagule. Accordingly, the low diffusion glassy phase may be a
feature of other realistic Turing system in the dilute limit, where
the discrete nature of the reactants becomes important.

\acknowledgements The authors thank  David Kessler and Philip Maini
for helpful discussion and comments. This work was supported by the
Israeli Science Foundation, grant no. 281/03 and by Yeshaya Horowitz
Fellowship..


\begin{thebibliography}{}
\bibitem{turing}A. M. Turing, Phil. Trans. Roy. Soc. \textbf{B 237} 37 (1952).
\bibitem{murray} See, e.g. J.D. Murray, \emph{Mathematical Biology}
(Springer-Verlag, New-York,  1993).
\bibitem{chem} V. Castets, E. Dulos, J. Boissonade and P. De Kepper, Phys.
Rev. Lett., \textbf{64}, 2953 (1990).
\bibitem{cross} M. C. Cross and P. Hohenberg, Rev.  Mod.
Phys. {\bf 65}, 851 (1993); P.K. Maini et. al., J. Chem. Soc.
Faraday T rans. \textbf{93}, 3601 (1997).
\bibitem{meron+} J. B. Wilson and A. D.Q. Agnew, Adv. Ecol. Res. \textbf{23}, 263
(1992); R. Lefever and O. Lejeune, Bull. Math. Biol. \textbf{59},
263 (1997); J. von Hardenberg, E. Meron, S. Shachak, and Y. Zarmi,
Phys. Rev. Lett. {\bf 87}, 198101 (2001).
\bibitem{lavee} N. M. Shnerb, P. Sarah, H. Lavee, and S. Solomon, Phys. Rev. Lett. {\bf 90}, 038101 (2003).
\bibitem{clm} K. Kaneko, Physica (Amsterdam) 34D, 1
(1989); J. Kockelkoren, A. Lemaitre and H. Chate, Physica \textbf{A
288}, 326 (2000).
\bibitem{lattice} Pattern formation and
Turing instability  have been recently demonstrated for a
Belousov-Zhabotinsky system dispersed in water droplets of a reverse
AOT microemulsion, see V. K. Vanag and I.R. Epstein, Phys. Rev.
Lett. \textbf{87}, 228301 (2001). The reaction takes place on
diffusivly coupled, discrete patches; one may expect a transition to
the glassy phase if the typical  time associated with the migration
of the slow reactant molecule between neighboring droplets is large
enough.
\bibitem{rsa}See, e.g., J. Talbot, G. Tarjus, P. R.Van-Tassel, and P.Viot,
Colloids Surf. A, {\bf 165}, 287 (2000), and references therein.
\end{thebibliography}
\end{document}